\documentstyle[aps]{revtex}
\begin{document}

\title{
Particle hopping models and traffic flow theory\\
}

\author{%
Kai Nagel\\
{}~\\
TSA-DO/SA, MS M997, Los Alamos National Laboratory, NM~87545, U.S.A.,
kai@lanl.gov\\
and\\
Santa Fe Institute, 1399 Hyde Park Rd., Santa Fe, NM~87501, U.S.A.,
kai@santafe.edu\\
{}~\\
This version: \today
}
\maketitle

\begin{abstract}
This paper shows how particle hopping models fit into the context of
traffic flow theory.  Connections between fluid-dynamical traffic flow
models, which derive from the Navier-Stokes-equations, and particle
hopping models are shown.  In some cases, these connections are exact
and have long been established, but have never been viewed in the
context of traffic theory.  In other cases, critical behavior of
traffic jam clusters can be compared to instabilities in the partial
differential equations.  Finally, it is shown how this leads to a
consistent picture of traffic jam dynamics.
\end{abstract}

\section{Introduction}

Traffic jams have always been annoying. At least in the industrialized
countries, the standard reaction has been to expand the transportation
infrastructure to match demand.  In this phase of fast growth,
relatively rough planning tools were sufficient.  However, in the last
years most industrialized societies started to see the limits of such
growth.  In densely populated areas, there is only limited space
available for extensions of the transportation system; and we face
increasing pollution and growing accident frequencies as the downsides
of mobility.  In consequence, planning is now turning to a fine-tuning
of the existing systems, without major extensions of facilities.  This
is for example reflected in the United States by the Clean Air Act and
by the ISTEA (Intermodal Surface Transportation and Efficiency Act)
legislation.  The former sets standards of air quality for urban
areas, whereas the latter forces planning authorities to evaluate land
use policies, intermodal connectivity, and enhanced transit service
when planning transportation.

In consequence, planning and prediction tools with a much higher
reliability than in the past are necessary.  Due to the high
complexity of the problems, analytical approaches are infeasible.
Current approaches are simulation-based (e.g.~\cite{ TRANSIMS,
Verkehrsverbund.NRW, Wylie.PARAMICS, TRAF}), which is driven by
necessity, but largely enhanced by the widespread availability of
computing power nowadays.  Yet, also for computers one needs good
simplified models of the phenomena of interest: Just coding a perfect
representation of reality into the computer is not possible because of
limits of knowledge, limits of human resources for coding all these
details, and because of limits of computational resources.

Practical simulation has to observe trade-offs between {\em
resolution}, {\em fidelity}, and {\em scale}~\cite{ Barrett.personal}.
Resolution refers to the smallest entities (objects, particles,
processes) resolved in a simulation, whereas fidelity means the degree
of realism in modeling each of these entities, and scale means the
(spatial, temporal, $\ldots$) size of the problem.  It is empirically
well known, for example from fluid dynamics, that to a certain extent
a low fidelity high resolution model (lattice gas automata~\cite{
Frisch.Hasslacher.Pomeau, Stauffer.CA}) can do as well as a high
fidelity low resolution model (discretization of the
Navier-Stokes-equations), or in short: Resolution can replace
fidelity.

Current state-of-the-art traffic modeling has a fixed unit of
(minimal) resolution, and that is the individual traveler.  Since one
is aiming for rather large scales (for example the Los Angeles area
consists of approx.\ 10~million potential travelers), it is rather
obvious that one has to sacrifice fidelity to achieve reasonable
computing times.

One important part of transportation modeling is road traffic.  For
example in Germany, road traffic currently contributes more than 81\%
of all passenger and 52.7\% of all freight transportation~\cite{
Verkehr.in.Zahlen.93}.  And despite widespread efforts, the share of
road transportation is still increasing.  For that reason, it makes
sense to start with road traffic when dealing with transportation
systems.

Putting these arguments together, one thing which is needed for large
scale transportation simulations is a {\em minimal\/} representation
of road traffic.  Particle hopping models clearly are candidates for
this, and even if not, building a {\em minimal\/} theory of road
traffic is certainly the right starting point.

This paper shows how particle hopping models fit into the context of
traffic flow theory.  It starts out with a historical overview of
traffic flow theory (Section~\ref{historic}), followed by a systematic
review of fluid-dynamical models for traffic flow (Section~\ref{fdyn})
starting from the Navier-Stokes-equations.  Section~\ref{defs} defines
different particle hopping models which are of interest in the context
of traffic flow.  Section~\ref{comparison} then shows the different
connections between the fluid-dynamical traffic flow models and
particle hopping models.  In some cases, these connections are exact
and have long been established, but have never been viewed in the
context of traffic theory.  In other cases, critical behavior of
traffic jam clusters can be compared to instabilities in the partial
differential equations.  Finally, it is shown how this leads to a
consistent picture of traffic jam dynamics~(Section~\ref{beyond}).  A
discussion of the consequences for traffic simulations
(Section~\ref{consequences}) serves as summary and discussion, and a
collection of open questions (Section~\ref{open}) conclude the paper.

\section{Historical overview}

\label{historic}

Vehicular traffic has been a widely and thoroughly researched area in
the 1950s and 60s.
For a review of traffic theory, see, for
example, one of~\cite{ Gazis, Gerlough, May}.

Vehicular traffic theory can be broadly separated into two branches:
Traffic {\em flow\/} theory, and {\em car-following\/} theory.

Traffic flow theory is concerned with finding relations between the
three fundamental variables of traffic flow, which are velocity~$v$,
density~$\rho$, and current or throughput or flow $j$.  Only two of
these variables are independent since they are related through $j =
\rho \, v$.

The first task of traffic flow theory historically was to search for
time-{\em in\/}dependent relations between $j$, $\rho$ and $v$, the
so-called fundamental diagrams.  The form of such a relation is,
though, still debated in the traffic flow literature~\cite{
Agyem.Hall, Hall.mea1}.  The problem stems mainly from the fact that
reality measurements are done in non-stationary conditions.  There,
only short time averages make sense, and they usually show large
fluctuations.  I will at the end of the paper discuss how a dynamic,
particle based description of traffic can account for these
difficulties.

The second step of traffic flow theory was to introduce a dynamic,
i.e.\ time-dependent description.  This was achieved by a well-known
paper from Lighthill and Whitham, published in 1955~\cite{
Lighthill.Whitham}.  This paper introduces a description based on the
equation of continuity plus the assumption that flow (or velocity)
depend on the density only, i.e.\ there is no relaxation time
(instantaneous adaption).

Prigogine, Herman, and coworkers developed a kinetic theory for
traffic flow~\cite{ Prigogine.book}.  They derived the
Lighthill-Whitham situation as a limiting case of the kinetic theory.
The kinetic theory anticipates many of the phenomena which come up in
later work, but probably because the mathematics of working in this
framework is fairly laborious, this theory has not been developed any
further until recently~\cite{ Helbing.kinetic}.

Instead, in 1979, Payne replaced the assumption of instantaneous
adaption in the Lighthill-Whitham theory by an equation for inertia,
which is similar to a Navier-Stokes-equation~\cite{ Payne}.  K\"uhne,
in 1984, added a viscosity term and initiated using the methods of
nonlinear dynamics for analyzing the equations~\cite{ Kuehne,
Kuehne.deutsch, Sick, Roediger}.

In a parallel development, Musha and Higuchi proposed the noisy
Burgers equation as a model for traffic and backed that up by
measurements of the power spectrum of traffic count data~\cite{
Musha.etc}.

In the next section, these fluid-dynamical models will be put into a
common perspective.

Car-following theory regards traffic from a more microscopic point of
view: The behavior of each vehicle is modeled in relation to the
vehicle ahead.  As the definition indicates, this theory concentrates
on single lane situations where a driver reacts to the movements of
the vehicle ahead of him.  Many car-following models are of the form
\begin{equation}
a(t+T) \propto {v(t)^m \over [\Delta x(t)]^l} \cdot \Delta v(t) \ ,
\end{equation}
where $a$ and $v$ are the acceleration and velocity, respectively, of
the car under consideration, $\Delta x$ is the distance to the car
ahead, $\Delta v$ is the velocity difference to that car, and $m$ and
$l$ are constants.  $T$ is a delay time between stimulus and response,
which summarizes all delay effects such as human reaction time or time
the car mechanics needs to react to input.

Other examples for car-following equations are $v(t+T) \propto \Delta
x$~\cite{ Newell.car.foll, Whitham.car.foll} or $a(t) \propto V[\Delta
x(t)] - v(t)$~\cite{ Bando.jpn.car.foll}, where $V[\Delta x]$ gives a
preferred velocity as a function of distance headway.

Mathematically, parts of this theory are very similar to the treatment
of atomic movements in crystals, and give results about the stability
of chains of cars (``platoons'') in follow-the-leader
situations.

One of the achievements of traffic theory of that period was that
relations between car-following models and {\em static\/}
flow-density-relations were derived.

Car-following theory will not be treated any further in this paper.

A more recent addition to the development of vehicular traffic flow
theory are cellular automata (CA).  Actually, the first proposition of
such a model is from Gerlough in 1956~\cite{ Gerlough.CA} and has been
further extended by Cremer and coworkers~\cite{ Cremer.Ludwig,
Schuett}.  They implemented fairly sophisticated driving rules and
also used single-bit-coding with the goal to make the simulation fast
enough to be useful for real-time traffic applications.  The bit-coded
implementation, though, made it too impractical for many traffic
applications.

In 1992, CA models for traffic were brought into the statistical
physics community.  Biham and coworkers used a model with maximum
velocity one for one- and for two-dimensional traffic~\cite{
Biham.etc}.  One-dimensional here refers to roads etc., and includes
multi-lane traffic.  Two-dimensional traffic in the CA context usually
means traffic on a 2-d grid, as a model for traffic in urban areas.
Nagel and Schreckenberg introduced a model with maximum velocity five
for one-dimensional traffic, which compared favorably with real world
data~\cite{ Nagel.92.Schreckenberg}.  Both approaches were further
analyzed and extended in a series of subsequent papers, both for the
one-dimensional~\cite{
Chung.1d.CA, Nagatani.1.lane, Yukawa.traff.CA,
%
Nagel.93.Herrmann, Vilar,
Nagatani.2.lane,
Takayasu.Takayasu,
Csahok.Vicsek,
Baldus,
Csanyi.Kertesz,
Emmerich.Rank,
Nagel.94.ltimes,
Nagel.94.Rasmussen,
Nagel.94.Schleicher,
Nagel.94.TRB,
Nagel.95.Paczuski,
Nagel.95.thesis,
Schadschn.Schrecki,
Schrecki.etc,
Ben.Naim, Migowsky.Rujan,
Latour,
Rickert.94.thesis,
Rickert.95.etc} (see also~\cite{ Blue}) and
the two-dimensional (see, e.g., \cite{ Martinez.2d.CA,
Nagatani.2.level, Freund.Poeschel}) investigations.

The motivation here is twofold.  The primary motivation is again the
aiming for computational speed, but this time to make statistical
analysis possible.  The second motivation is that the model is still
simple enough to be treated analytically, but distinctively different
from other particle-hopping models.  In addition, CA-methodology is
planned to be used as a high-speed option in traffic projects in
Germany~\cite{ Verkehrsverbund.NRW} and in the United States~\cite{
TRANSIMS}.

{}From a theoretical point of view, the methodology of CA is placed
between fluid-dynamical and car-following theories, and is helpful to
further clarify the connections between these approaches.  This paper
aims at contributing to the first part, i.e.\ understanding and
clarifying the relations between particle-hopping models and
fluid-dynamical models for traffic flow.

\section{Fluid-dynamical models for traffic flow}

\label{fdyn}

This section reviews fluid-dynamical models for traffic flow.  The
models can broadly be distinguished into two classes: Models which do
not include the effects of inertia and models which do.  Models of the
first class can be derived from the equation of continuity where
velocity or current are defined as functions of the density only.
Models of the second class are formally Navier-Stokes-equations, with
a car-specific force term which takes into account that drivers want
to drive at a certain desired speed.  If one sets the time constant of
this force term to zero, i.e.\ assuming {\em instantaneous adaption\/}
to the surrounding density, then one recovers the models of the first
class.

\subsection{General equations}

Papers on traffic flow theory usually start with stating the equations
under consideration, without setting them in perspective.  I will
therefore in this paper attempt a more fundamental approach, similar
to conventional fluid-dynamics.  The precise presentation of most of
these equations is necessary anyhow because the particle-hopping
models presented later relate to these equations.

One might use the standard fluid-dynamical conservation equations for
mass and momentum as a starting point for a fluid-dynamical
description of traffic:
\begin{equation}
\partial_t \rho + \partial_x ( \rho \, v ) = 0
\label{ eq.cont}
\end{equation}
and
\begin{equation}
{\hbox{d} v \over \hbox{d} t} \equiv \partial_t v + v \cdot \partial_x v
= F/m \ ,
\label{ eq.mom}
\end{equation}
where $\rho$ is the density and $v$ the velocity.  $\hbox{d} / \hbox{d} t$ is
the
individual (Lagrangian)
derivative, $F$ is the force acting on mass $m$.  The first equation
(of continuity) describes mass conservation; the second one (momentum
conservation) describes the fact that the momentum of a point of mass
may only be changed by a force.  Obviously, for traffic, $F$ has to
include vehicle and driving dynamics.

\subsection{Fluctuations}

A standard first step in fluid-dynamics \cite{ Cotton.clouds} is to
assume that $v$ and $\rho$ fluctuate statistically around average
values $\langle v \rangle$ and $\langle \rho \rangle$, i.e.
\begin{equation}
v = \langle v \rangle + v' \ , \qquad \langle v' \rangle = 0
\end{equation}
and
\begin{equation}
\rho = \langle \rho \rangle + \rho' \ , \qquad \langle \rho' \rangle = 0 \ .
\end{equation}
In this case, one only assumes that $\langle v \rangle$ and $\langle
\rho \rangle$ fluctuate {\em slowly\/} in space and time; for the
general subtleties of hydro-dynamical theory see, e.g.,~\cite{
Landau.hydro}.  Inserting these relations into (\ref{ eq.cont}) and
(\ref{ eq.mom}) and subsequent averaging over the whole equations
(e.g.\ $ \left\langle \partial_x [(\langle \rho \rangle+\rho')\,(
\langle v \rangle +v')] \right\rangle
= \partial_x \langle \rho \rangle \langle v \rangle
+ \partial_x \langle \rho' v' \rangle
$)
yields
\begin{equation}
\partial_t \langle \rho \rangle
+ \partial_x \langle \rho \rangle \langle v \rangle
+ \partial_x \langle \rho' v' \rangle = 0
\end{equation}
and
\begin{equation}
\partial_t \langle v \rangle_L + \langle v \rangle_L \partial_x \langle v
\rangle_L + {1 \over 2} \partial_x \langle v' v' \rangle
= \langle F/m \rangle \ .
\end{equation}

One often parameterizes averaged fluctuations by the corresponding
gradient~\cite{ Cotton.clouds} $ \langle v' A' \rangle \approx - \alpha \,
\partial_x \langle A \rangle $, which leads to the set of equations
\begin{equation}\hbox{\vtop{\halign{&$\displaystyle#$\hfil\cr
\partial_t \rho + \partial_x ( \rho \, v ) & = D \, \partial_x^2 \rho \cr
\partial_t v + v \, \partial_x v & = \nu \, \partial_x^2 v + F/m \ , \cr
}}}
\label{ n.s.eqs}
\end{equation}
where, according to convention, the averaging brackets have been
omitted, and the diffusion coefficient $D$ as well as the (kinematic)
viscosity $\nu$ are assumed to be independent of $x$ and $t$.  It
should be noted that similar diffusion terms can also be obtained from
other arguments.

\subsection{Lighthill-Whitham-theory and kinematic waves}

If one assumes that the velocity is a function of density only ($v =
f(\rho)$), then the momentum equation is no longer necessary.  This
corresponds to instantaneous adaption; the particles (or cars) carry
no memory.  Using without loss of generality the current $j(\rho)
\equiv \rho \, v(\rho)$, and setting in addition $D=0$, from (\ref{
n.s.eqs}) one obtains
\begin{equation}
\partial_t \rho + j'(\rho) \, \partial_x \rho = 0
\end{equation}
(Lighthill-Whitham-equation~\cite{ Lighthill.Whitham}), where $j' = \hbox{d}
j / \hbox{d} \rho$.  $j'$ will, in this paper, always mean the derivative of
$j$, even when the prime in connection with other letters can denote
fluctuations.  For a review of this theory, see, e.g.,~\cite{
Lighthill.Whitham, Haberman}.

The equation can be solved by the ansatz $\rho(x,t) = \rho(x-ct)$
with
\begin{equation}
c = j'(\rho) \ .
\end{equation}
This allows the solution of the characteristics (see, e.g., \cite{
Haberman}): A region with density $\rho$ travels with constant
velocity $c=j'(\rho)$, and the resulting straight line in space-time
is called characteristic.  When $j(\rho)$ is convex, i.e.\ $j''<0$,
then for regions of decreasing density ($\rho(x_1) > \rho(x_2)$ for
$x_1 < x_2$) the characteristics separate from each other.  In regions
of {\em in\/}creasing density, the characteristics come closer and
closer together.  When two characteristics touch each other, a density
discontinuity appears at this place (a front), which moves with
velocity
\begin{equation}
c = { j(x_2) - j(x_1) \over \rho(x_2) - \rho(x_1) }
= { \Delta j \over \Delta \rho } \ .
\end{equation}
Note that formally the fluid-dynamical description has broken down
here because both $\rho$ and $j$ are no longer continuous functions of
$x$.

An illustrative example is a queue, such as at a red light.  When the
light turns green, the outflow front quickly smoothes out, whereas the
inflow front remains steep.

Note that usually at maximum flow $c=j'=0$.  Structures which operate
at maximum flow do not move in space.

Leibig~\cite{ Leibig} gives results of how a random initial
distribution of density steps in a closed system evolves towards two
single steps according to the Lighthill-Whitham-theory.

\subsection{Lighthill-Whitham with dissipation}

Adding dissipation to the Lighthill-Whitham-equation leads to
\begin{equation}
\partial_t \rho + j'(\rho) \, \partial_x \rho = D \, \partial_x^2 \rho
\ .
\label{ l.w.diss}
\end{equation}
The solution of this equation is again a non-dispersive wave with phase
and group velocity $j'$.
The difference is that $D$ introduces dissipation (damping) of the
wave: The amplitude decays as $e^{-D k^2}$, where $k$ is the
wavenumber.  This reflects the intuitively reasonable effect that
traffic jams should tend to dissolve under homogeneous and stationary
conditions.

\subsection{The nonlinear diffusion (Burgers) equation}

For a further development, $j(\rho)$ has to be specified.  Since we
are mostly interested in the behavior of traffic near maximum
throughput, we start by choosing the simplest mathematical form which
yields a ``well-behaved'' maximum:
\begin{equation}
j(\rho) = v_{max} \, \rho ( 1 - \rho ) \ ,
\label{ Greenshields}
\end{equation}
which, in traffic science, is called the
Greenshields-model~(see~\cite{ Gerlough}).  $v_{max}$ is, in
principle, a free parameter, but it has an interpretation as the
maximum average velocity for $\rho \to 0$.
Mathematicians would set $v_{max}=1$; traffic scientists use $1 -
\rho/\rho_{\hbox{jam}}$ for the term in parenthesis.  $\rho_{\hbox{jam}}$
is the density of vehicles in a jam.
%
The maximum current $j_{max}$ is reached at $\rho(j_{max}) = 1/2$.

Substituting (\ref{ Greenshields}) into (\ref{ l.w.diss}) yields
\begin{equation}
\partial_t \rho + v_{max} \, \partial_x \rho - 2 v_{max} \, \rho \, \partial_x
\rho
= D \, \partial_x^2 \rho \ .
\end{equation}
Musha and coworkers~\cite{ Musha.etc} have shown that by introducing a
linear transformation of variables
\begin{equation}
x = v_{max} \, t' - x' \ , \qquad t = t' \ ,
\end{equation}
one obtains
\begin{equation}
\partial_{t'} \rho + 2 \, v_{max} \, \rho \, \partial_{x'} \rho
= D \, \partial_{x'}^2 \rho \ ,
\end{equation}
which is the (deterministic) Burgers equation.

This equation has been investigated in great detail by Burgers~\cite{
Burgers} as the simplest non-linear diffusion equation.  The
stationary solution is a uniform density $\rho(x,t) = const$.  A
single disturbance from this state evolves over time into a
characteristic triangular structure with amplitude $\sim t^{-1/2}$,
width $\sim t^{1/2}$, bent to the right such that the right side of
the disturbance becomes discontinuous, and moving to the right with
velocity $c = j' = 2 \, \rho \, v_{max}$.

When interpreting this for traffic jams, one has to re-transform the
coordinates.  Jams can then move {\em both\/} to the left or to the
right (with velocities between $v_{max}$ and $-v_{max}$), and the
discontinuous front develops at the inflow side of the jam, i.e.\
where the vehicles enter the jam.  One sees that this solution is just
the solution of the characteristics, with a dissipating diffusion term
added---as should be expected because of $D>0$.

Some other versions of the Burgers equation are relevant for traffic
and have been investigated thoroughly~\cite{ Krug.Spohn, Binder.etc,
Majumdar.Barma.tags}:

{\bf Noisy Burgers equation:} Adding a Gaussian noise term
$\eta$ to the equation (i.e.~$\langle \eta(x,t) \eta(x',t') \rangle = \eta_0
\, \delta(x-x') \, \delta(t-t')$) leads to the noisy Burgers equation
\begin{equation}
\partial_t \rho + 2 \, v_{max} \, \rho \partial_x \rho = D \partial_x^2 \rho +
\eta \ .
\end{equation}
This equation does no longer reach a stationary state.

{\bf Generalized Burgers equation:} The nonlinearity of the
Burgers equation can be generalized:
\begin{equation}
\partial_t \rho = \sum_\beta b_\beta \partial_x \rho^\beta + D \partial_x^2
\rho \ .
\end{equation}

Generalized Burgers equations with arbitrary $\beta$ have been
investigated~\cite{ Binder.etc, Krug.Spohn}.

\subsection{Including momentum}

The equations so far do not explain the spontaneous phase separation
into relatively free and rather dense regions of vehicles which is
observed in real traffic. To obtain this, one has to include the
effect of momentum: One can neither accelerate instantaneously to a
desired speed nor slow down without delay.  It becomes necessary to
include the momentum equation.  Here, one has to specify the force
term $F/m$, which describes acceleration and slowing down.  At least
two properties are usually incorporated, which are called the
``relaxation term'' and the ``interaction term''.

A first order approximation for the relaxation term is~\cite{
Kuehne, Payne}
\begin{equation}
{1 \over \tau} ( V(\rho) - v ) ,
\end{equation}
where $V(\rho)$ is the desired average speed as a function of density,
and $\tau$ is a relaxation time.  This choice yields exponential
relaxation towards the desired speed.  The function $V(\rho)$ has to
be specified externally, for example from measurements.

A commonly used interaction term~\cite{ Cremer, Kerner.Konh,
Kuehne, Payne} is
\begin{equation}
- { c_0^2 \over \rho } \partial_x \rho \ . \label{interaction}
\end{equation}
The meaning is that one tends to reduce speed when the density
increases, even when the local density is still consistent with the
current speed.

$c_0$ is treated as constant; in traffic, a typical value for $c_0$ is
$15$~km/h~\cite{ Kuehne.deutsch}.

Formally, this term derives from the pressure term of compressible
flow, $-(1/\rho) \partial_x p$, where $p$ is the pressure due to
thermal motion of the particles.  Assuming an ideal gas ($p = \rho \,
R \, T$) and isothermic behavior $T=const$, one obtains waves similar
to sound waves as a solution of the linearized
equations.\footnote{True sound waves, though, would assume the gas to
behave adiabatic, i.e.\ $p\propto\rho^\kappa$.} This leads to
(\ref{interaction}), where $c_0$ is the speed of the ``sound''
waves. (See below for a short discussion.)

Note that sound waves move in {\em both\/} directions from a
disturbance, which means that sound waves alone are not a good
explanation for freeway start-stop-waves, contrary to what is
sometimes written~\cite{ Begley.newsweek}.

A possible momentum equation for traffic therefore is~\cite{ Kuehne}
\begin{equation}
\partial_t v + v \partial_x v = - { c_0^2 \over \rho } \partial_x \rho
+ {1 \over \tau} \, [V(\rho) - v] + \nu \partial_x^2 v \ .
\label{KKK-momentum}
\end{equation}
Since one now has two variables, one also needs an equation of
continuity to close the system:
\begin{equation}
\partial_t \rho + \partial_x ( \rho \, v ) = D \, \partial_x^2 \rho \ .
\label{KKK-continuity}
\end{equation}
Usually, $D$ is set to zero.

For this equation, the homogeneous solution $(v,\rho) \equiv
(v_0,\rho_0)$ is unstable for densities near maximum flow for a
suitable choice of parameters.  Using the methods of nonlinear
dynamics, K\"uhne and coworkers~\cite{ Kuehne, Roediger, Sick} went
beyond linear stability analysis (see also \cite{ Hilliges,
Kurtze.Hong}).  One finds a multitude of stable
or unstable fixpoints and limit cycles which suggest that traffic near
maximum flow operates on a strange attractor.  This can lead to
quasi-periodic behavior, exactly as is observed in traffic
measurements.

Earlier work~\cite{ Cremer, Payne} has analyzed the same
equation without viscosity ($\nu = 0$).

\subsection{Discussion of fluid-dynamical approaches}

Fluid-dynamical models have been used in traffic science for a long
time, with considerable success.  But they have shortcomings.  Some of
the major points are:

(i)~One has to give externally the relation between speed or current
and density.  This is unsatisfying in terms of the development of a
theory.  But an even more intricate problem is that there is no
agreement on a functional form of the speed-density relation; it is
even under discussion if this relation is at all continuous~\cite{
Hall.mea1, Persaud.Hall.phase.trans}.

(ii)~Temperature parameterizes the random fluctuations of particles
around their mean speed.  For gases, fluctuations and therefore
temperature increase with density. For granular media, fluctuations
{\em de\/}crease with density (i.e.\ inside a jam) -- it has been
claimed that exactly this inverse temperature effect is responsible
for clustering~\cite{ Goldhirsch.temperature.gran.mat}.  In this way,
assuming isothermic instead of adiabatic behavior as done for the
momentum equation seems only half the way one has to go.
Helbing~\cite{ Helbing} discusses this further.

(iii)~Helbing~\cite{ Helbing} also discusses the effect of excluded
volume to take into account the spatial extension of vehicles.

Nonetheless, fluid-dynamical approaches~\cite{ Kerner.Konh, Kuehne,
Roediger, Sick} give, for the first time, systematic insight into
traffic dynamics near maximum flow beyond simple extrapolation of
light and dense traffic results.  These results will be further
discussed near the end of this paper.

\section{Definitions of particle hopping models}

\label{defs}

This section defines several particle hopping models which are
candidate models for traffic.  They all have in common that they are
defined on a lattice of, say, length~$L$, where $L$ is the number of
sites, and that each site can be either empty, or occupied by exactly
one particle.  Also, in all models particles can only move in one
direction.  The number of particles, $N$, is conserved except at the
boundaries.

\subsection{The Stochastic Traffic Cellular Automaton (STCA)}

The Stochastic Traffic Cellular Automaton (STCA), which has been
treated in a series of papers~\cite{
%
Csahok.Vicsek,
Baldus,
Csanyi.Kertesz,
Emmerich.Rank,
Nagel.94.ltimes,
Nagel.94.Rasmussen,
Nagel.94.Schleicher,
Nagel.94.TRB,
Nagel.95.thesis,
Nagel.95.Paczuski,
Schadschn.Schrecki,
Schrecki.etc%
}, is defined as follows.  Each particle ($=$ car) can
have an integer velocity between 0 and $v_{max}$.  The complete
configuration at time-step~$t$ is stored, and the configuration at
time-step~$t+1$ is computed from that (parallel or synchronous
update).  For each particle, the following steps are done in
parallel:\begin{itemize}

\item
Find number of empty sites ahead ($= gap$) at time~$t$.

\item
If $v > gap$ ({\bf too fast}), then slow down to $v:=gap$. [rule~1]

\item
Else if ($v<gap$) ({\bf enough headway}) {\em and\/} $v<v_{max}$, then
accelerate by one: $v:=v+1$.  [rule~2]

\item
{\bf Randomization:} If after the above steps the velocity is larger
than zero ($v > 0$), then, with probability $p$, reduce $v$ by
one. [rule~3]

\item
{\bf Particle propagation:} Each particle moves $v$ sites
ahead. [rule~4]

\end{itemize}
The randomization condenses three different properties of human
driving into one computational operation: Fluctuations at maximum
speed, over-reactions at braking, and retarded (noisy) acceleration.

Note that, because of integer arithmetic, conditions like $v>gap$ and
$v \ge gap+1$ are equivalent.

When the maximum velocity of this model is set to one ($v_{max}=1$),
then the model becomes much simpler: For each particle, do in
parallel:\begin{itemize}

\item
If site ahead was free at time~$t$, move, {\em with
probability~$1-p$}, to that site.

\end{itemize}
Since the STCA shows different behavior for $v_{max} \ge 2$ than for
$v_{max}=1$, I will distinguish them using STCA/1 and STCA/2,
respectively.

\subsection{The cruise control limit of the STCA (STCA-CC)}

In the so-called cruise control limit of the STCA~\cite{
Nagel.95.Paczuski}, fluctuations at free driving, i.e.\ at maximum
speed and undisturbed by other cars, are set to zero.
Algorithmically, this amounts to the following: For all cars do in
parallel:\begin{itemize}

\item
A vehicle is stationary when it travels at maximum velocity~$v_{max}$
and has free headway: $gap \ge v_{max}$.  Such a vehicle just
maintains its velocity.

\item
Else (i.e.\ if a vehicle is not stationary) the standard rules of the
STCA are applied.

\end{itemize}
Both acceleration and braking still have a stochastic component.  The
stochastic component of braking is realistic, but it is irrelevant for
the results presented here.

\subsection{The deterministic limit of the STCA (CA-184)}

One can take the deterministic limit of the STCA by setting the
randomization probability~$p$ equal to zero, which just amounts to
skipping the randomization step.  It turns out that, when using a
maximum velocity $v_{max}=1$, this is equivalent~\cite{ Krug.Spohn} to
the cellular automaton rule 184 in Wolfram's notation~\cite{ Wolfram},
which is why I will use the notation CA-184/1 and CA-184/2.

Much work using CA models for traffic is based on this model.  Biham
and coworkers~\cite{ Biham.etc} have introduced it for traffic flow,
with $v_{max}=1$.  Other authors base further results on it~\cite{
Chung.1d.CA, Nagatani.1.lane, Yukawa.traff.CA, Nagel.93.Herrmann, Vilar,
Takayasu.Takayasu,
Csahok.Vicsek}.  Some~\cite{ Nagel.93.Herrmann, Vilar} also use it
with $v_{max}$ larger than one.  It is also the basis of the
two-dimensional CA models for traffic~\cite{ Martinez.2d.CA,
Nagatani.2.level, Freund.Poeschel}.

\subsection{The cruise control version for the CA-184 (CA-184-CC)}

Takayasu and Takayasu~\cite{ Takayasu.Takayasu} introduced a model
which amounts to a deterministic cruise control situation for
CA-184/1.  This may not be obvious from the rules, but it will become
clear from the dynamic behavior summarized later.  Since they use only
maximum velocity $v_{max}=1$, the rules are short: For all particles
do in parallel:\begin{itemize}

\item
A particle with velocity one is moved one site ahead when the site
ahead is free ($gap \ge 1$).

\item
A particle at rest ($v=0$) can only move when $gap \ge 2$.

\end{itemize}
Generalizations to maximum velocity larger than one are
straightforward, but do not seem to lead to additional insight.

\subsection{The Asymmetric Stochastic Exclusion Process (ASEP)}

The probably most-investigated particle hopping model is the
Asymmetric Stochastic Exclusion Process (ASEP).  It is defined as
follows:\begin{itemize}

\item
Pick one particle randomly. [rule~1]

\item
If the site to the right is free, move the particle to that
site. [rule~2]

\end{itemize}

The ASEP is closely related to CA-184/1 and STCA/1 (i.e.\ both with
maximum velocity one).  The difference actually only is in the update
schemes, not in the rules which are used for each particle.

In contrast, in CA-184/1, one picks {\em all\/} particles
synchronously and moves them according to rule~2 of the ASEP. In order
to make this work, one has to use ``old'' information (i.e.\ from
iteration $t$) to decide if the site to the right is free (i.e.\ if
the particle can be there at time $t+1$).  For the ASEP, this
distinction between ``old'' and ``new'' information is not necessary
because one only picks one particle at a time and all others do not
move.

In STCA/1 (with $p=1/2$), one picks randomly {\em half of all\/}
particles synchronously and moves them according to rule~2 of the
ASEP.

In order to compare the ASEP with the other, synchronously updated
models, one has to note that, in the ASEP, {\em on average\/} each
particle is updated once after $N$ single-particle updates.  A
time-step (also called update-step or iteration) in the ASEP is
therefore completed after $N$ single-particle updates ($=N$ attempted
hops).

It was already noted earlier~\cite{ Krug.Spohn} that going from ASEP
to CA-184/1, i.e.\ changing the update from asynchronous to
synchronous, produces very different dynamics. In this paper, I will
in addition show that re-introducing the randomness via the
randomization (rule~4) in the STCA again leads to different results.

A systematic way of reducing the noise for the ASEP could be done
using techniques described by Wolf and Kertesz~\cite{
Wolf.noise.reduction}, i.e.\ by putting a counter on each particle and
move it only after $k$ trials.  For large $k$ it becomes more and more
improbably that one particle is moved twice while a neighboring
particle is not moved at all during that time.  Taking the limit $k
\to \infty$ then reduces the ASEP to the CA-184 process in a smooth
way.

One can also define higher velocities for the ASEP by simply replacing
ASEP--rule~2 by STCA/2--rules~1, 2, and~4.  In such a case, each
particle has to remember its velocity $v$ from the last move.

\section{Particle hopping models, fluid dynamics, and critical
exponents}

\label{comparison}

Both for the ASEP/1 and for the CA-184/1, fluid-dynamical limits and
critical exponents are well known (see, e.g., \cite{
Majumdar.Barma.tags, Krug.Spohn, Janowsky.burgers, Binder.etc}).  The
most straightforward way to put the concept of critical exponents into
the context of traffic flow is to consider ``disturbances'' (i.e.\
jams) of length $x$ and ask for the time $t$ to dissolve them.  For
example, one would intuitively assume that a queue of length~$x$ at a
traffic light which just turned green would need a time $t$
proportional to $x$ until everybody is in full motion.  By this
argument, the dynamic exponent~$z$, defined by $t \sim x^z$, should be
one.

Yet, there may be more complicated cases.  Imagine again a queue at a
traffic light just turned green but this time also some fairly high
inflow at the end of the queue.  The jam-queue itself will start
moving backwards, clearing its initial position in time $t \sim x$.
However, the dissolving of the jam itself may be governed by different
rules.  An example for this will be given in the following.

\subsection{ASEP/1}

It can be shown that the classic stochastic asymmetric exclusion
process corresponds to the noisy Burgers equation (see, e.g.,~\cite{
Krug.Spohn, Binder.etc}).  More precisely, the particle process
corresponds to a diffusion equation $ \partial_t \rho + \partial_x j =
D \, \partial_x^2 \rho + \eta $ with a current~\cite{ Krug.Spohn,
Derrida.etc} of $j = \rho \, (1-\rho)$.  Inserting this yields
\begin{equation}
\partial_t \rho + \partial_x \rho - \partial_x \rho^2
= D \, \partial_x^2 \rho + \eta \ ,
\label{Burgers-traffic}
\end{equation}
which is exactly the Lighthill-Whitham-Greenshields case with noise
and diffusion described earlier.  {\em In other words, the ASEP/1
particle hopping process and the Lighthill-Whitham-theory (plus noise
plus diffusion), specialized to the case of the Greenshields
flow-density relation, describe the same behavior.}

In the steady state, this model shows kinematic waves ($=$ small
jams), which are produced by the noise and damped by diffusion
(Fig.~\ref{3-V1-rs}). These non-dispersive waves move forwards (wave
velocity $c = j' = 1-2\rho > 0$) for $\rho < 1/2$ and backwards
($c<0$) for $\rho>1/2$ (Fig.~\ref{5-V5-rs}). At $\rho=1/2$, the wave
velocity is exactly zero ($c=0$), and this is the point of maximum
throughput~\cite{ Krug}.  If traffic were modeled by the ASEP, then
one could detect maximum traffic flow by standing on a bridge:
Jam-waves moving in flow direction indicate too low density (cf.\
Fig.~\ref{3-V1-rs}), jam-waves moving against the flow direction
indicate too high density.

The ASEP is one of the cases where clearing a site follows a different
exponent than dissolving a disturbance.\footnote{%
Note that technically, all these remarks are only valid for {\em
small\/} disturbances.  The problem is that if one is no longer close
to the steady state, one sees transient behavior which may be
different~\protect\cite{ Krug.Spohn}.
} As long as $\rho \ne 1/2$, a disturbance of size~$x$ moves with
speed $c \ne 0$ and therefore clears the initial site in time $t \sim
c \cdot x \sim x^1$, i.e.\ with dynamical exponent $z=1$.  In order to
see how the disturbance itself dissolves, one transforms into the
coordinate system of the wave velocity.  One conventionally does that
by first separating between the average density~$\langle\rho\rangle_L$
and the fluctuations~$\rho'$. By inserting $\rho=\langle\rho\rangle_L
+ \rho'$ one obtains
\begin{equation}
\partial_t \rho' + (1 - 2 \langle\rho\rangle_L) \, \partial_x \rho'
- 2 \, \rho' \, \partial_x \rho' = D \, \partial_x^2 \rho' + \eta \ .
\end{equation}
When transforming this into the moving coordinate system $x' = x +
(1-2\langle\rho\rangle_L) \cdot t$, one obtains
\[
\partial_t \rho' - 2 \, \rho' \, \partial_x' \rho'
= D \, \partial_x'^2 \rho' + \eta \ ,
\]
which is the classic noisy Burgers equation.

For this equation it is well known that the dynamical exponent is
$z=3/2$ (``KPZ'' exponent).  In other words, in the original
coordinate system a disturbance four times as big as another one,
$x'=4x$, needs $t' \sim x' = 4x \sim 4t$, i.e.\ four times as much
time to clear the site, but $t' \sim x'^{3/2} = (4x)^{3/2} \sim 8t$,
i.e.\ 8~times as much time until the jam-structure itself is no longer
visible in the noise. A precise treatment of this uses, e.g.,
correlations between tagged particles~\cite{ Majumdar.Barma.tags}.

The drawback of this model with respect to traffic flow is that it
does neither have a regime of laminar flow nor have ``real'', big jams.
Because of the random sequential update, vehicles with average
speed~$\overline{v}$ fluctuate severely around their average position
given by $\overline{v} \, t$.  As a result, they always ``collide''
with their neighbors, even at very low densities, leading to
``mini-jams'' everywhere.  This is clearly unrealistic for light
traffic.

Actually, this fact is also visible in the speed-density-diagram.
Using the Greenshields flow-density relation, one obtains
\begin{equation}
v = { j \over \rho } \propto 1-\rho \ .
\end{equation}
This is in contrast to the observed result that, at low densities,
speed is nearly independent of density (practically no interaction
between vehicles).

\subsection{ASEP/2}

Judging from space-time plots, changing the maximum velocity in the
update from $v_{max}=1$ to $v_{max} \ge 2$ does not change the
universality class~\cite{ Schrecki.etc} (see Figs.~\ref{3-V1-rs}
and~\ref{5-V5-rs}). It skews the flow-density-relation towards lower
densities, but does not lead to other phenomenological behavior.

\subsection{CA-184}

Using a maximum velocity higher than one does not change the general
behavior of CA-184.  It therefore makes sense to directly discuss the
general case.

As explained above, the CA-184/1 is the deterministic counterpart of
the ASEP/1.  But taking away the noise from the particle update
completely changes the universality class (i.e.\ the exponent
$z$)~\cite{ Krug.Spohn}.  The model now corresponds to the
non-diffusive, non-noisy equation of continuity
\begin{equation}
\partial_t \rho + j' \, \partial_x \rho = 0
\label{ 184/2-fdyn}
\end{equation}
with a (except at $\rho=\rho_{jmax}$) linear flow
\begin{equation}
j' = {\hbox{d}j \over \hbox{d}\rho} = \cases{
v_{max} & for $\rho<\rho_{jmax}$ \cr
-1 & for $\rho > \rho_{jmax}$. \cr
}
\label{ 184/2-fdiag}
\end{equation}

The intersection point of the fundamental diagram divides two
phenomenological regimes: light traffic ($\rho < \rho_{jmax}$) and
dense traffic ($\rho > \rho_{jmax}$).

A typical situation for light traffic is shown in Fig.~\ref{light.tty}
(with $v_{max}=5$).  After starting from a random initial condition,
the traffic relaxes to a steady state, where the whole pattern just
moves $v_{max}=5$ positions to the right in each iteration.  Cars
clearly have a tendency of keeping a $gap$ of $\ge v_{max}=5$ between
each other.  As a result, the current, $j$, in this regime is
\begin{equation}
j_< = \rho \cdot v_{max} \ .
\end{equation}
The velocity of the kinematic waves in this regime is $c_< = j'_< =
v_{max}$.  This means that disturbances, such as holes, just move with
the traffic, as can also be seen in Fig.~\ref{light.tty}.

Dense traffic is different (Fig.~\ref{dense.tty}).  Again starting
from a random initial configuration, the simulation relaxes to a
steady state where the whole pattern moves one position to the {\em
left\/} in each iteration.  Note that cars still move to the right; if
one follows the trajectory of one individual vehicle, for this car
regions of relatively free movement are alternating with regions of
high density and slow speed.  Although in a too static way, this
captures some of the features of start-stop-traffic.  The average
speed in the steady state equals the number of empty sites divided by
the number of particles: $\langle v \rangle_L = (L-N) / N$; the
current is $j_> = \rho \cdot \langle v \rangle_L$, or, with $\rho =
N/L$,
\begin{equation}
j_> = 1-\rho \ .
\end{equation}
This straight line intersects with the one from light traffic at $\rho
= 1 / (1 + v_{max})$, which is therefore the density corresponding to
maximum throughput $j_{max} = v_{max} / (1+v_{max})$.

The velocity of the kinematic waves in the dense regime is $j'_> =
-1$, which corresponds to the backwards moving pattern in
Fig.~\ref{dense.tty}.

Since the second term of Eq.~\ref{ 184/2-fdyn} (with~\ref{
184/2-fdiag}) is (except at $\rho=\rho_{jmax}$) linear in the
density, these are linear Burgers equations, and the dynamic exponent
$z$ is equal to 1~\cite{ Krug.Spohn}.

More precisely, the following happens: The outflow of a jam in this
model always operates at flow $j_{out}=j_{max}$ and density
$\rho_{out}=\rho_{jmax}$.  The time $t$ until a jam of length~$x$
dissolves therefore obeys the average relation $t \propto x / (j_{max}
- j_{in})$, where $j_{in}$ is the average inflow to the jam.  Since
$j\propto\rho$ for $\rho \le \rho_{jmax}$, one can
write that as
\begin{equation}
t \propto {x \over \rho_{jmax} - \rho(j_{in})} \ .
\label{det.scaling}
\end{equation}
This means that for $\rho < \rho_{jmax}$, the
critical exponent~$z$ is indeed one, but at $\rho = \rho_{jmax}$, $t$
diverges.  This effect is also visible when disturbing the system from
its stationary state: The transient time $t_{trans}$ until the system
is again stationary scales as~\cite{ Nagel.93.Herrmann}
\begin{equation}
t_{trans} \sim {1 \over \rho_{jmax} - \rho } \ .
\label{det.transient}
\end{equation}

The scaling law~(\ref{det.transient}) is actually also true for
$\rho>\rho_{jmax}$, albeit for a different reason
with a slightly more complicated phenomenology.  See~\cite{
Nagel.93.Herrmann} for more details.

Two observations are important at this point:

(i)~Many papers in the physics literature~\cite{ Biham.etc,
Chung.1d.CA, Nagatani.1.lane, Yukawa.traff.CA, Nagel.93.Herrmann, Vilar,
Nagatani.2.lane, Csahok.Vicsek} use this model for their
investigations.  Also the 2d-grid models (see, e.g.,~\cite{
Martinez.2d.CA, Nagatani.2.level, Freund.Poeschel}) essentially use
this model for the one-dimensional part of their movements, although
the two-dimensional interactions seem to change the flow-density
relationship~\cite{ Molera.Cuesta}.  The CA-184 model lacks at least
two features which are, as I will argue later, important with respect
to reality: (a)~Bi-stability: Laminar flow above a certain density
becomes instable, but can exist for long times. (b)~Stochasticity:
CA-184 is completely deterministic, i.e.\ a certain initial condition
always leads to the same dynamics.  Real traffic, however, is
stochastic, that is, even identical initial conditions will lead to
different outcomes, and a model should be capable of calculating some
distribution of outcomes (by using different random seeds).

(ii)~The fluid-dynamical model behind the so-called cell transmission
model~\cite{ Daganzo.cell.transmission}, which is a discretization of
the Lighthill-Whitham-theory, is similar to Eq.~\ref{ 184/2-fdyn}
with~\ref{ 184/2-fdiag}, especially with respect to the range of
physical phenomena which are represented.  The only difference is that
the $j$-$\rho$-relation of Ref.~\cite{ Daganzo.cell.transmission} has
a flat portion at maximum flow instead of the single peak of Eq.~\ref{
184/2-fdiag}.  That means that in the cell transmission model low
density and high density traffic behave similarly to CA-184, but
traffic at capacity has a regime where waves do not move at all.

Using other $j$-$\rho$-relations in discretized
Lighthill-Whitham-models (e.g.~\cite{ Mahm.DYNASMART, Hilliges}), will
lead to other relations for the wave speeds, but the range of physical
phenomena (backwards or forwards moving waves) which can be
represented will always resemble CA-184; especially, neither the
bi-stability nor the stochasticity can be represented.

\subsection{CA-184-CC}

No fluid-dynamical limits for the other particle hopping models are
known.  Yet, results for the jam dynamics for the cruise control
situations~\cite{ Takayasu.Takayasu, Nagel.95.Paczuski} offer valuable
insights, which will be described in the following.

The important new feature of the cruise control version of CA-184 is a
bi-stability~\cite{ Takayasu.Takayasu}.  Using $v_{max}=1$ in this
section ($v_{max} > 1$ does not seem to offer additional insight),
this bi-stability occurs between two densities, i.e.\ for $\rho_{c1} <
\langle\rho\rangle_L < \rho_{c2} = 1/2$, where $\langle\rho\rangle_L
:= N/L$, and, for $v_{max}=1$, $\rho_{c1}=1/3$ and $\rho_{c2}=1/2$.
$\langle . \rangle_L$ means the average over the whole (closed) system
of length~$L$.  In this range, some initial conditions lead to laminar
flow but others lead to traffic including jams.  Takayasu and Takayasu
found the following:

(i)~Starting with maximally spaced particles and initial velocity one,
one finds stable configurations with flow $\langle j \rangle_L =
\langle\rho\rangle_L \cdot v_{max} = \langle\rho\rangle_L$ for low
densities $\langle\rho\rangle_L \le 1/2 =: \rho_{c2}$.  For high
densities $\langle\rho\rangle_L > \rho_{c2}$, a jam phase appears for
{\em all\/} initial conditions since not all particles can keep $gap
\ge 1$.  Once a jam has been created, all particles in the outflow of
this jam have $gap = 2$.  For $t \to \infty$, this dynamics
reorganizes the system into jammed regions with density one and zero
current, and laminar outflow regions with $\rho_{out}=1/3$ and
$j_{out}=1/3$.  Simple geometric arguments then lead, for the whole
system, to $\langle j \rangle_L = (1-\langle\rho\rangle_L)/2$ and
$\langle v \rangle_L = (1/\langle\rho\rangle_L-1)/2$.

(ii)~Starting, however, with an initial condition where all particles
are clustered in a jam, this jam is only sorted out up to
$\langle\rho\rangle_L \le 1/3 =: \rho_{c1}$, leading to $\langle j
\rangle_L=\langle\rho\rangle_L$ and $\langle v \rangle_L=1$.  For
$\langle\rho\rangle_L > \rho_{c1}$, the initial jam survives forever,
yielding $\langle j \rangle_L=(1-\langle\rho\rangle_L)/2$ and $\langle
v \rangle_L=(1/\langle\rho\rangle_L-1)/2$.  One observes that, for
$\rho_{c1} < \langle\rho\rangle_L < \rho_{c2}$, this initial condition
leads to a different final flow state than the initial conditions
in~(i). --- Note that $\rho_{c1}$ is equal to the outflow density
$\rho_{out}$.

(iii)~Starting from an arbitrary initial condition, the
density-velocity relation converges to one of the above two types.

Note that up to before this section, all relations between $j$, $v$,
and $\rho$ were also locally correct, which is why averaging brackets
were omitted.  Now, this is no longer true.  For example densities
slightly above $\rho_{c2}$ do not really exist on a local level; they
are only possible as a global composition of regions with local
densities $\rho = \rho_{c1}$ plus others with local densities
$\rho=1$.

Since the model is deterministic, one can calculate the behavior from
the initial conditions.  For any particle~$i$ with initial velocity
zero one can determine the influence that particle has on following
particles~$i+1,i+2,\ldots$.  For particle $i+k$ to be the first one
not to be involved in the jam caused by $i$, one needs the average gap
between $i$ and $k$ to be larger than two.  This corresponds to a
density between $i$ and $k$ of $\rho_{ik} < 1 / (gap+1) = 1/3 =
\rho_{c1}$.  The sequence $(gap_{i+j})_j$ describes a random walk,
which is positively (negatively) biased for $\rho > \rho_{c1}$ ($\rho
< \rho_{c1}$), and unbiased at the critical point $\rho = \rho_c =
\rho_{c1}$~\cite{ Takayasu.Takayasu}.

\subsection{STCA-CC/1}

The cruise control limit of the STCA is in some sense a mixture
between the CA-184 and the full STCA.  Since the STCA-CC has no
fluctuations at free driving, the maximum flow one can reach is with
all cars at maximum speed and $gap=v_{max}$.  Therefore, one {\em
can\/} manually achieve flows which follow, for $\rho \le \rho_{c2}$,
the same $j$-$\rho$-relationship as the CA-184, where $\rho_{c2}$ now
denotes the density of maximum flow of the deterministic model CA-184,
i.e.\ $\rho_{c2} = 1/(v_{max}+1)$.

Above a certain $\rho_{c1}$, these flows are unstable to small local
perturbations.  This density will turn out to be a ``critical''
density; for that reason I will use $\rho_c \equiv \rho_{c1}$. Many
different choices for the local perturbation give rise to the same
large scale behavior.  The perturbed car eventually re-accelerates to
maximum velocity.  In the meantime, though, a following car may have
come too close to the disturbed car and has to slow down.  This
initiates a chain reaction -- an emergent traffic jam.

It is straightforward to see~\cite{ Nagel.95.Paczuski} that $n(t)$,
the number of cars in the jam, follows a usually biased, absorbing
random walk, where $n(t)=0$ is the absorbing state (jam
dissolved): Every time a new car arrives at the end of the jam, $n(t)$
increases by one, and this happens with probability $j_{in}$, which is
the inflow rate.  Every time a car leaves the jam at the outflow side,
$n(t)$ decreases by one, and this happens with probability $j_{out}$.
When $j_{in}=j_{out}$, $n(t)$ follows an unbiased absorbing random walk.
$j_{in} \ne j_{out}$ introduces a bias or drift term $\propto (j_{in}
- j_{out}) \cdot t$.

This picture is consistent with Takayasu and Takayasu's observations
for the CA-184-CC model.  The main difference is that now {\em both\/}
the inflow gaps and the outflow gaps form a random sequence.  Another
difference is conceptually: Takayasu and Takayasu have looked at the
transient time starting from initial conditions, whereas Nagel and
Paczuski look at jams starting from {\em a single disturbance}.  The
latter leads to a cleaner picture of the traffic jam dynamics because
it concentrates on the transition from laminar to start-stop-traffic
which is observed in real traffic.

The statistics of such absorbing random walks can be calculated
exactly. For the unbiased case one finds that
\begin{equation}
\langle n(t) \rangle \sim t^\eta \ ,
\qquad P_{surv}(t) \sim t^{-\delta}
\hbox{ and}
\qquad \langle w(t) \rangle_{surv} \sim t^{\eta+\delta} \ ,
\end{equation}
where $P_{surv}$ is the survival probability of a jam until time~$t$,
$w(t)$ means the width of the jam, i.e.\ the distance between the
leftmost and the rightmost car in the jam.  $\langle . \rangle$ means
the ensemble average over {\em all\/} jams which have been initiated,
and $\langle . \rangle_{surv}$ means the ensemble average over
{\em surviving\/} jams.  For the critical exponents, one finds as well
from theory as from numerical simulations $\delta=1/2$ and $\eta=0$.

$\eta=0$ re-confirms that, at the critical density~$\rho_c$, jams in
the average {\em barely survive\/} (unbiased random walk).

If one now uses $j_{in}$ as order parameter, and, say, $P_{surv}(t)$
as control parameter, then we have a second order phase transition,
where
\begin{equation}
P_{surv}(t) \cases{%
= 0 & for $j_{in} < j_{out}$ and $t \to \infty$,\cr
\sim t^{-\delta} & for $j_{in} = j_{out}$ and $t \to \infty$, and\cr
= const & for $j_{in} > j_{out}$ and $t \to \infty$.\cr
}
\end{equation}
For that reason, we call $j_c := j_{out}$ the critical flow, and the
associated density $\rho_c := \rho(j_c)$ the critical density.

It is important to note that $j_{in} > j_{out}$ as a stable, longtime
state is only possible due to the particular definition of the cruise
control limit and in an open system.  If one would use a closed
system, the outflow of the jam would eventually go around the loop and
turn into the inflow of the jam (see Fig.~\ref{V1-CC}), leading to the
situation $j_{in} = j_{out}$; if one would go away from the cruise
control limit, eventually other jams would form upstream of the one
under consideration, and the outflow of these jams would eventually be
the inflow of the jam under consideration, again leading to $j_{in} =
j_{out}$.

Deviations from the cruise control limit will be addressed later; let
us now consider a {\em closed\/} system (periodic boundary conditions,
i.e.\ traffic in a loop):\begin{itemize}

\item
For $\rho < \rho_c$ and arbitrary initial conditions, jams are
ultimately sorted out.  Then, every car has velocity $v = v_{max}$ and
$gap \ge v_{max}$, is thus in the free driving regime as defined above.

\item
For $\rho_c \le \rho \le \rho_{c2}$, the long-time behavior depends on
the initial conditions.  For example, even in the extreme case of
$\rho = \rho_{c2}$, the state where every car has velocity $v =
v_{max}$ and $gap = v_{max}$ is stable and results in a flow of $j =
v_{max} \cdot \rho_{c2}$.  However, most other initial configurations will
lead to jams, and for the limit of infinite system size, at least one
of them never sorts out.

\item
For $\rho > \rho_{c2}$, all initial conditions lead to jams.

\end{itemize}
Note that this is again consistent with the results of Takayasu and
Takayasu for the CA-184-CC system~\cite{ Takayasu.Takayasu}.

\subsection{STCA-CC/2}

Replacing maximum velocity $v_{max}=1$ by $v_{max} \ge 2$ does not
change the critical behavior, but it adds a complication~\cite{
Nagel.95.Paczuski}.  Now, jam clusters can branch, with large jam-free
holes in between branches of the jam.  As a result, space-time plots
of such jams now appear to show fractal properties, and in simulations
at the critical density~$w(t)$ does not follow any longer a clean
scaling law, whereas $n(t)$ and $P_{surv}$ still do.

The explanation of this is that the holes in the jam are large enough
to cause logarithmic corrections to the width, but not large enough to
make it completely fractal.  More precisely, the hole size
distribution $P_h(x)$, i.e.\ the probability to find a hole of
size~$x$ in a given equal time cut ($=$ jam configuration) scales as
\begin{equation}
P_h(x) \sim x^{-\tau_h} \ ,
\end{equation}
where both from a theoretical argument and from simulations
$\tau_h=2$.  Yet, it is known that for $\tau_h \le 2$ the fractal
dimension for such a configuration is $D_f = \tau_h-1$ (see, e.g.,
\cite{ Maslov.95.P.B}).  In this sense, such a traffic
jam cluster operates at the ``edge of fractality''.

And such a hole size distribution causes logarithmic corrections to
the width when $n(t)$ is given:
$
\langle w(t) \rangle_{surv}
\sim \langle n(t) \rangle_{surv} \, ( 1 + c \, \log t ) \ .
$

\subsection{STCA/1}

For the STCA at $v_{max}=1$, from visual inspection
(Fig.~\ref{STCA-1}) one finds that distinguishable jams do not exist
here.  Instead, the space-time plot looks much more like one from the
ASEP.

This is confirmed by theoretical analysis.  Schreckenberg and
coworkers~\cite{ Schadschn.Schrecki, Schrecki.etc} have performed
analytical calculations for the stationary state throughput~$j$ given
$\rho$ in a closed system using $n$-point correlations ($n$-cluster
method) and found that for $v_{max}=1$ this analysis is already exact
for $n=2$.  This is no longer true for higher $v_{max}$.  For the
ASEP, for the same analysis, the mean-field approximation, i.e.\
$n=1$, is already exact.  The difference between the ASEP and the
STCA/1 in this analysis is that in the STCA/1 one finds an effective
repulsive force of range one between particles, caused by the parallel
update.  This helps to keep particles more equidistant than in the
ASEP case, thus leading to a higher flow.

\subsection{STCA/2}

For $v_{max} \ge 2$, the $n$-cluster analysis does no longer lead to
an exact solution, indicating that a different dynamical regime has
now completely taken over.  (In practice, though, the $n$-cluster
analysis is already fairly close to simulation results for $n \simeq
5$.)  Visual inspection of space-time plots confirms that the dynamics
now is much more similar to the cruise control limit, i.e.\ to
STCA-CC/2, than to the ASEP.

Yet, in contrast to the cruise control limit, here multiple jams exist
simultaneously.  Jams start spontaneously and independently of other
jams because vehicles fluctuate even at maximum speed, as determined
by a parameter $p_{free} \neq 0$.

The STCA displays a scaling regime near the density of maximum
throughput $\rho_{jmax}$, but with an upper
cutoff at $t \simeq 10^4$ which was observed to depend on
$p_{free}$~\cite{ Nagel.94.ltimes}.  One can attribute this cutoff to
the non-separation of the time scales between disturbances and the
emergent traffic jams~\cite{ Nagel.95.Paczuski}.  As soon as
$p_{free}$ is different from zero, the spontaneous initiation of a new
jam can terminate another one.  Obviously, this happens more often
when $p_{free}$ is high, which explains why the scaling region gets
longer when one reduces $p_{free}$.

Cs\'anyi and Kert\'esz~\cite{ Csanyi.Kertesz} find long-range
connectivity of jams in the STCA/2 (using $p=1/4$) only at densities
much higher than $\rho_{jmax}$.  Yet, it is
still unclear how this is related to the the percolation picture
described in~\cite{ Nagel.94.ltimes} and~\cite{ Nagel.95.Paczuski},
and if, for example, going away from the cruise control limit and
changing $p$ to a value much different from 1/2 will separate the
remains of a percolation transition point from the onset of long-range
connectivity.

A helpful analogy is droplet formation in a gas-liquid
transition~\cite{ Lifschitz.kinetik}, where gas corresponds to the
laminar phase and the droplets correspond to the jams.  The gas always
``tests'' (in fluctuations) simultaneously at many positions if
droplets can survive.  When one neglects surface tension, then
droplets can{\em not\/} survive at sub-critical density, they {\em
can\/} survive at supercritical density, and they {\em barely\/}
survive exactly at the critical density, making macroscopic
fluctuations maximal at this point.  --- Note that neglecting the
surface tension of the droplets changes the nature of the phase
transition from first order to second order.

\section{Going beyond: Traffic jam dynamics}

\label{beyond}

All these results together put us into a position to draw a fairly
consistent picture of traffic jam dynamics.

\subsection{An intuitive starting point}

Measurements of human driving behavior show that over a fairly large
velocity range, $gap$ is proportional to velocity. $gap$ here is
$\Delta x - L$, where $\Delta x$ is the front-bumper-to-front-bumper
distance ($=$ distance headway) between two cars, and $L$ is the
length one car occupies (in the average) in a jam.  This makes
intuitively a lot of sense, since it reflects the fact that the {\em
time\/} gap should be approximately the same as the delay time~$T$
which is needed between seeing the brake lights and actually starting
to brake, and should therefore be largely independent of velocity
(Pipes' theory, cf.~\cite{ May}).\footnote{Note that traffic science
traditionally does not include some ``security space'' into the
definition of $L$~\protect\cite{ May}; therefore ``gap'' and ``time
gap'' are somewhat different here.}

Field studies (cf.~\cite{ May}) indeed confirm that the delay time is
approximately constant for velocities between 15~and 40~miles per hour
(between 24~and 64~km/h, data from the 1950s). This delay time
consists of several components, including, e.g., reaction time or the
time needed for actually pressing the brake pedal, and it is of the
order of one second.

Therefore, we have $gap = T \cdot v$. Using the average relations
$1/\Delta x = \rho$ and $1/L = \rho_{jam}$ (density inside a jam), we
obtain
\begin{equation}
v = {1 \over T} \left( {1 \over \rho} - {1 \over \rho_{jam}} \right) \ .
\end{equation}
Thus, for high density, for the current $j$ one has
\begin{equation}
j_{high} = \rho \, v = {1 \over T} \left( 1 - {\rho \over \rho_{jam}} \right) \
{}.
\end{equation}

For low density, one can assume that there is some (average) $v_{max}$
which is independent of $gap$ for large enough spacings, and
therefore, for low densities
\begin{equation}
j_{low} = \rho \, v_{max} \ .
\end{equation}

At $j_{c2}$ and $\rho_{c2}$ , these two curves intersect and thereby
define the maximum flow according to this model.  Assuming, say,
$v_{max} = 120$~km/h, $T=1.1$~sec, and $L=7.5$~m, one obtains
$\rho_{c2} \approx 1/45$~m and $j_{c2} \approx 2650$~vehicles per hour
per lane, which is slightly above the highest 5-minute-averages which
are obtained in reality (e.g.~\cite{ Hall.mea1, Hall.Pushkar}).

This model is essentially equivalent to the CA-184/2 particle hopping
model.

As a side remark, note that traffic security experts teach drivers
that one should reach the position of the car ahead only after more
than two seconds, which is independent of velocity and reflects the
fact that time headway is approximately equal to time gap.  But it is
interesting to see that this would actually lead to a maximum current
of 2~cars per second or 1800~cars per hour, much less than the up to
2400~cars per hour which are observed.

\subsection{More realistic traffic jam dynamics}

Yet, real traffic behaves differently.  At high densities, we do not
observe the homogeneous velocity $v = gap/T$ as predicted by the
intuitive argument above, but relatively free flow which is
inter-dispersed by start-stop-waves.  This is confirmed by measurements
of the $j$-$\rho$-relation, where, instead of lining up on a single
curve, the measurements form a fairly scattered data cloud especially
in the region of the flow maximum.

The explanation of this is as follows.  Laminar traffic will always
and at all densities, due to small fluctuations, have small
disturbances which can develop into jams.  The inflow to the jam
decides if such a jam can become long-lived or not: Since the average
outflow $j_{out}$ is fixed by the driving dynamics, $j_{in} > j_{out}$
makes the jam (in the average) long-lived, $j_{in} < j_{out}$ not.
$j_{in} = j_{out}$ defines a critical point, i.e.\ $j_c \equiv j_{out}$ and
$\rho_c \equiv \rho_{out}$, where traffic jam clusters in the average
barely survive, as e.g.\ quantified by $\langle n(t) \rangle = t^\eta$
with $\eta = 0$.

All this is true for an open system, or a system which is large enough
and where times are short enough so that the closed boundaries are not
felt.  In a closed system, the jams ultimately absorb all the excess
density $\rho_{\hbox{\footnotesize\it excess}} = \rho - \rho_c$; as a
result, all traffic between jams operates at $\rho_c$.

Average measurements of the long time behavior of traffic flow can
therefore show at most $j_c = j_{out}$.

Again, the gas-liquid analogy (without surface tension) is helpful.
Gas can also be brought into a super-cooled regime, for example by
increasing the density while keeping the temperature constant.  But
this state is only meta-stable, and eventually droplets will form and,
in a closed system, absorb excess density until the density
surrounding the droplets is exactly at the critical point (without
surface tension!).

This dynamical picture explains the high variations in the short time
measurements.  Measuring at a fixed position in a situation like in
Fig.~\ref{spontaneous}, one can measure arbitrary combinations of
supercritical laminar traffic, critical laminar traffic, jams, or
traffic during acceleration or slowing down.  See Fig.~\ref{fdiags}
for a comparison between short-time (300~time steps) averages and a
schematic picture.  Data points along the (a) branch belong to stable
and laminar traffic.  Data points along the (c) branch belong to still
laminar, but only meta-stable traffic.  Data points along the (d)
branch belong to creeping high density traffic.

All other data points are mixtures between regimes, where two or more
regimes have been captured during the 300~iterations interval.
Essentially, these data points should lie between point (b) and branch
(d), yet, due to high fluctuations and due to the effects of
acceleration and braking, which are not captured in the steady state
arguments, we see huge fluctuations.  For example, when a car is just
leaving a jam, the density decreases, but the velocity adaption is
lagging somewhat behind.  Therefore, the car has too low speed for the
given density, leading to too low a flow value.

This picture also makes precise the hysteresis argument of Treiterer
and coworkers~\cite{ Treiterer.hysteresis}, also confirmed
later~\cite{ Persaud.Hall.phase.trans, Hall.Pushkar}.  These
measurements confirm the idea that the traffic density can go above
the critical point while still being laminar, similar to the gas which
can be super-cooled by increasing the density.  Yet, both for traffic
and for super-cooled gases, this state is only meta-stable and
eventually leads to a phase separation into jams and laminar flow.
Quantitative evidence of this will be given in a separate paper (in
preparation).

The picture is also consistent with recent results both in
fluid-dynamical models and mathematical car-following models for
traffic flow.  In Ref.~\cite{ Kerner.Konh}, traffic simulations using
a fluid-dynamical model starting from nearly homogeneous conditions
eventually form stable waves.  The fluid-dynamical model one can, by
usual linearization, find the parameters for the onset of instability.
What is $\langle n(t) \rangle \sim t^\eta$ in the particle hopping
model becomes the amplitude $A(t) \sim e^{\lambda t}$ in the
fluid-dynamical model, and at the onset of instability $A = const$ is
similar to $\langle n(t) \rangle = const$ (since $\eta = 0$).
Therefore, the wave in the fluid-dynamical model corresponds to the
{\em average\/} jam-cluster in the particle-hopping model.

Lee~\cite{ Lee.waves} explains the underlying mechanism for a model
for granular media.  He distinguishes ``dynamic'' from ``kinematic''
waves.  Dynamic waves are found in the K\"uhne/Kerner/Konh\"auser
equations (Eqs.~\ref{KKK-momentum} and~\ref{KKK-continuity})
when the relaxation time $\tau > 0$; they are similar to sound waves
in gases.  Kinematic waves are found in the same equations when $\tau
\to 0$, in which case the equations reduce to the Lighthill-Whitham
case.  The wave formation mechanism thus is that the instability first
triggers the ``sound'' wave.  The density inside the wave increases
and outside the wave decreases until both densities are outside the
instable range.  Then the kinematic mode takes over.

Kurtze and Hong~\cite{ Kurtze.Hong} make this more precise for the
K\"uhne/Kerner/Konh\"auser equations: Below the critical density, the
kinematic wave with wave velocity $c = j' = \hbox{d} j / \hbox{d}
\rho$ is the only solution of the linearized equation, and this
solution is stable.  At the critical density, this solution bifurcates
into two unstable solutions with wave velocity $c = j' \pm \epsilon$,
where $\epsilon \to 0$ for $\rho \searrow \rho_c$, and $\epsilon$ is
the equivalent of the speed of sound.

Bando and co-workers~\cite{ Bando.jpn.car.foll} also find the
separation of traffic into laminar and jammed phases in a
deterministic continuous mathematical car-following model.

\section{Consequences for traffic simulation models}

\label{consequences}

These findings have some fairly far-reaching implications for traffic
simulation models.

First, particle hopping models, which seem at the first glance as too
rough an approximation of reality, include the same range of dynamic
phenomena as the most advanced fluid-dynamical models for traffic flow
to date.  Yet, particle hopping models offer some distinctive
advantages for practical simulations.  Particle hopping models are
known to be numerically robust especially in complex geometries, and
realistic road networks with all their interconnections etc.\
certainly qualify as such.  Practical road network implementations of
the fluid-dynamical theory are so far only using the Lighthill-Whitham
equations, which are (without diffusion) marginally stable and can
certainly be made stable by using a stable numerical discretization
scheme.

Second, the fact that traffic jam behavior follows critical exponents
leads to the expectation that {\em all\/} microscopic models which
include jam formation should display the same critical exponents: The
universality hypothesis of critical phenomena states that critical
exponents are fairly robust against changes in microscopic rules.  For
the traffic jam case, this is backed up by the fact that the exponents
can be theoretically explained.  The consequence for traffic
simulation is that, as long as one expects certain simple aspects of
traffic jam formation to be realistic enough for the problem under
consideration, e.g.\ for large scale questions, {\em the simplest
possible model will be sufficient for the task}, thus saving human and
computational resources.

Third, the present results show that close-up car-following behavior
is not the most important part to model.  The important part is to
model {\em deviations\/} from the optimal (smooth) behavior and in
which way they lead to jam formation.  Another important part, which
seems far from obvious, is the {\em acceleration behavior}, especially
when there are other cars ahead, since it is the acceleration behavior
which mostly decides about maximum flow out of a jam (which may be a
simple traffic light!).  Note that this part of driving behavior is
much more connected to physical properties of the vehicles; electric
vehicles with their different acceleration capabilities might completely
change traffic flow~\cite{ Barrett.personal}.  Therefore,
investigations such as this paper {\em are\/} important for
microscopic modeling as long as one does not have the perfect model of
driving, or not the computational resources to run it.

Fourth, fast running and easy to implement particle hopping models can
be very useful in interpreting measurements.  Measurements such as for
the traditional 5-minutes-averaged fundamental diagrams (flow vs.\
density vs.\ velocity) have increasingly recognized the fact that the
dynamics around the measurement site has an extreme influence on the
outcome of the measurements, thus making the results far from
universal.  This point will be further discussed in another paper (in
preparation).

Fifth, particle hopping models are inherently microscopic, which
allows to add individual properties to each car such as identity of
travelers, route plan, engine temperature (for emission modeling) etc.
These properties are imperative for the kind of traffic models which
are needed in current policy evaluation processes.

And last but not least, particle hopping models are stochastic in
nature, thus producing different results when using different random
seeds even when starting from identical initial conditions.  At first,
this is certainly considered a disadvantage from the point of view of
policy makers or traffic engineers.  However, the traffic system is
inherently stochastic, and the variance of the outcomes is an
important variable itself.  How will we be able to distinguish
reliable from unreliable predictions without knowing something about
the range of possible outcomes? --- Furthermore, there is reason to
believe that the average over several stochastic runs will {\em not\/}
be identical to a deterministic run.  Imagine, for example, a case where
in a deterministic model, a queue at one intersection has a back-spill
which in the average {\em just\/} does not reach another
intersection.\footnote{By queue I mean a queue with spatial extension.
This is different from the use of the word in queuing theory.}  In
the stochastic model, the maximum length of this queue will, between
different simulation runs, fluctuate around its average value, thus
back-spilling into the other intersection in nearly 50\% of all
runs. Since this possibly disrupts traffic in this other intersection,
this can cause long-range effects and network breakdown.

\section{Some open questions}

\label{open}

Many open questions remain, though.  For example:

What is the exact relation between average cluster growth in CA
models, wave amplitude growth in fluid-dynamical models, droplet
growth in the liquid-gas transition interpretation, and phase space
portraits in car following models?

Is there a hydrodynamical limit for the STCA?  If so, how can it be
proven to be correct?  Do critical exponents help here?

What is the exact relation to granular media?  P\"oschel has both
observed and simulated similar waves for sand falling down in a narrow
tube~\cite{ Poeschel}.  He has also found in the simulations the
bi-stability leading to laminar flow or to jam waves depending on the
initial conditions.  Lee and coworkers~\cite{ Lee.Leibig, Lee.waves}
have related these waves to a fluid-dynamical theory which is similar
to the K\"uhne/Kerner/Konh\"auser theory for traffic flow.  Sch\"afer
finds a similar phase transition as the one stressed in this paper for
simulated granular flow, except that above the critical point, the
flow is exactly zero~\cite{ Schaefer.personal}; supposedly, such a
flow-density relation would also support the same overall dynamics.

What is the {\em minimal\/} ingredient for the instability which
causes the traffic breakdown?  Both the car-following models (CA and
continuous) and the fluid-dynamical approach have produced the
instability after adding an inertia term.  Yet, Goldhirsch and Zanetti
point out that an inverse temperature effect is responsible for the
clustering~\cite{ Goldhirsch.temperature.gran.mat}.

Can one say more about universality than in the last section?

What can one-dimensional theory say about two-dimensional problems,
such as they are regularly encountered for urban traffic problems?  A
series of papers (see, e.g., \cite{ Martinez.2d.CA, Nagatani.2.level,
Freund.Poeschel, Molera.Cuesta}) have used cellular automata
techniques for building models for town traffic.  These models use the
CA-184/1 model for driving dynamics, but add elements for directional
changes.  Molera and coworkers have built a theory for their
two-dimensional model~\cite{ Molera.Cuesta}, and their flow equation
is essentially a 2-dimensional version of the Lighthill-Whitham
equation {\em with a quadratic flow-density relation.}  That means
that adding stochastic directional changes would change the model from
CA-184 type to the ASEP type.

What is the relation to $1/f$-noise?  Musha and Higuchi have measured
$1/f$-noise in the power spectrum of a car detector time series~\cite{
Musha.etc}.  They explained this by a noisy Burgers equation, in a way
though which differs from Krug's interpretation~\cite{ Krug}. Nagel
and Paczuski~\cite{ Nagel.95.Paczuski} have predicted a precise $1/f$
law for the power spectrum of the density time series, which was
roughly confirmed by simulations for STCA-CC/2.  Yet, Nagel and
Herrmann find, using a continuous car-following model and following
the traffic movement, a $1/f^\alpha$ law, with $\alpha \approx
1.3$~\cite{ Nagel.95.Herrmann}. Car following is slightly different
from the particle hopping models in this paper; but if the arguments
in~\cite{ Nagel.95.Paczuski} were entirely correct, this should not
matter. --- Understanding $1/f$ noise behavior would be helpful
because it would be much easier to measure in reality than, say,
lifetime distributions~\cite{ Nagel.94.ltimes, Nagel.95.Paczuski}.

What is the meaning to the ongoing discussion about the value of
synchronous updating for explaining physical phenomena?  Huberman and
Glance~\cite{ Huberman.Glance.async} have re-issued the warning that
parallel updating may produce artifacts and that usually stochastic
asynchronous updating would be a better approximation of reality.
Yet, for the traffic case, it is clear from this paper that the
(synchronous) STCA produces a much better model for reality than the
(asynchronous) ASEP.  One would probably have to go to much higher
spatial and temporal resolutions (and thus loose all the computational
advantages) when one wanted to build a stochastically updating model
of traffic.

\section*{Acknowledgments}

I thank A.~Bachem, P.~Bak, C.~Barrett, S.~Esipov, J.~Lee, M.~Leibig,
H.J.~Herrmann, M.~Paczuski, S.~Rasmussen, M.~Rickert, J.~Sch\"afer,
M.~Schreckenberg, and D.E.~Wolf for discussions, hints and
encouragement.

A discussion group at TSA-DO/SA (LANL) about microscopic traffic
modeling, consisting of Chris Barrett, Steven Eubank, Steen
Rasmussen, Jay Riordan, Murray Wolinsky, and me, helped clarify many
issues.

Most of the ideas with respect to simulation are based on discussions
with Chris Barrett and Steen Rasmussen, reflecting work in progress
which is only to a small part published in~\cite{
Rasmussen.Barrett.simulation}.

I am especially grateful to S.~Rasmussen and M.~Rickert, who carefully
went through the last-to-final draft and helped me to prevent some
errors and make some issues more clear.

\begin{figure}
\caption{\label{3-V1-rs}%
Space-time plot for random sequential update,
$v_{max}=1$ (ASEP/1), and $\rho=0.3$.  Clearly, the kinematic waves are moving
forwards.  For $\rho>1/2$, the kinematic waves would be moving
backwards, and the plot would look similar to \protect\ref{5-V5-rs}.
}
\end{figure}

\begin{figure}
\caption{\label{5-V5-rs}%
Space-time plot for random sequential update, $v_{max}=5$ (ASEP/5),
and $\rho=0.3$, showing that higher maximum velocity does not lead to
a different appearance as long as one uses random sequential update.
}
\end{figure}

\begin{figure}
\caption{\label{light.tty}%
Space-time plot for CA-184/5 and subcritical density.
}
\end{figure}

\begin{figure}
\caption{\label{dense.tty}%
Space-time plot for CA-184/5 and supercritical density.
}
\end{figure}

\begin{figure}
\caption{\label{V1-CC}%
Space-time plot for STCA-CC/1, at supercritical density, with one
disturbance.  The jam first grows according to $n(t) \sim (j_{in} -
j_{out}) \cdot t$.  Eventually, via the periodic boundary conditions,
the outflow reaches the jam as inflow, and $n(t)$ follows a random
walk (apart from finite size effects).
}
\end{figure}

\begin{figure}
\caption{\label{09-V5-prll-cc}%
Space-time plot for parallel update (cruise control limit),
$v_{max}=5$, $\rho = 0.09$, i.e.\ slightly above critical.  The flow
is started in a deterministic, supercritical configuration, but from a
single disturbance separates into a jam and a region of exactly
critical density.---This is phenomenologically the same plot as
Fig.~\protect\ref{V1-CC} except that $v_{max}=5$.
}
\end{figure}

\begin{figure}
\caption{\label{STCA-1}%
Space-time plot for parallel update, $v_{max}=1$.
}
\end{figure}

\begin{figure}
\caption{\label{spontaneous}%
Space-time plot for parallel update, $v_{max}=5$,
$\rho = 0.09$ (i.e.\ slightly above $\rho(j_{max})$), starting from
ordered initial conditions.  The ordered state is meta-stable, i.e.\
``survives'' for about 300~iterations until is spontaneously separates
into jammed regions and into regions with $\rho =\rho(j_{max})$.
}
\end{figure}

\begin{figure}
\caption{\label{fdiags}%
Flow-density fundamental diagrams for the STCA.  {\em Top:\/}
Simulation output from the STCA.  Short-time averages are taken over
300 simulation steps and thus mimic the 5-minute averages often taken
in reality. {\em Bottom:\/} Schematical view.  (a) is the subcritical
branch, (b) is the critical point, (c) is the supercritical branch,
and (d) is the branch where traffic only creeps.  5-minute averages
at densities between $\rho_c$ at (b) and the creep branch are mixtures
between the dynamical regimes.
}
\end{figure}

\end{document}